\def\p{\partial}
\def\pl{\overleftarrow\p}
\def\pr{\overrightarrow\p}
\def\H{{\cal H}}

\def\F{{\cal F}}
\def\r{\rho}
\def\s{\sigma}
\def\o{\omega}
\def\Jxp{\langle\hat J\rangle_{xp}}
\hsize=12.8truecm\vsize=17.5truecm
\voffset=5truecm\hoffset=2truecm
\null\parindent=0pt
{\it to appear in: Proc. of the Oviedo Symposium 1996 on "New Developments
     on Fundamental Problems in Quantum Physics"; eds.: M.Ferrero 
     and A. van der Merwe, (Kluwer, 1997); also quant-ph/9610037}\hfill
\bigskip\bigskip 
\bf 
\leftline{COHERENT STATES AND THE MEASUREMENT PROBLEM}
\line{\ }\line{\ }\line{\ }\line{\ }\line{\ }
\line{\hskip 1.8truecm Lajos Di\'osi\hfill}
\bigskip
\it
\line
{\hskip 1.8truecm KFKI Research Institute for Particle and Nuclear Physics
\hfill}
\line
{\hskip 1.8truecm H-1525 Budapest 114, POB 49, Hungary\hfill}
\line{\ }\line{\ }
\rm
\parindent=0pt
The convenience of coherent state representation is discussed from the 
viewpoint of what is in a broad sense called the measurement problem in 
quantum mechanics. Standard quantum theory in coherent state representation 
is intrinsically related to a number of earlier concepts conciliating 
quantum and classical processes. From a natural statistical interpretation, 
free of collapses or measurements, the usual von Neumann-L\"uders collapse 
as well as its quantum state diffusion interpretation follow. 
In particular, a theory of coupled quantum and classical dynamics arises,
containing the fluctuation corrections versus the fenomenological
mean-field theories. 

\bigskip
Key words: 
quantum measurement, wave function collapse, open quantum systems.\hfill
\bigskip\bigskip
\parindent 1.3truecm

{\it 1. Introduction.\/}
The state of a canonical quantum system is characterized by its state vector
$\vert\psi\rangle$ satisfying the Schr\"odinger evolution equation.
The Schr\"odinger-equation, with Hamiltonian $\H(\hat x,\hat p)$, 
takes the equivalent form
$${d\over dt}\hat\r=-i\left[\H(\hat x,\hat p),\hat\r\right]
\eqno(1)$$
if we use the density operator $\hat\r=\vert\psi\rangle\langle\psi\vert$
to represent the system's quantum state.

In a certain sense, clarified ultimately by von Neumann [1], one may speak
about the probability distribution of the coordinate operator $\hat x$, 
which is given by the diagonal elements
$\langle x\vert\hat\r\vert x\rangle$ of the density operator
in coordinate representation.
This distribution contains no information about the
quantum system's momentum $\hat p$. {\it Vice versa}, in momentum 
representation the distribution $\langle p\vert\hat\r\vert p\rangle$ lacks 
information about the statistics of $\hat x$. 
Apparently, the proper statistical interpretation 
of the quantum state needs the distinguished notion of quantum measurement
and assumes abrupt changes, i.e. collapses, of the wave function [1,2]. 
Quantum theory becomes dichotomic: the evolution of the quantum state
is governed by the Schr\"odinger-equation (1) while in
quantum measurements the wave function performs random jumps.

Why couldn't one incorporate the statistical interpretation of the wave 
function into a suitable quantum dynamics free of collapses? 
In this talk I suggest that this can perhaps be done within
standard quantum mechanics merely by choosing a proper representation.

As the proper one, I suggest the well-known
coherent state representation $\vert x,p\rangle$ [3].
Then the diagonal elements
$\langle x,p\vert\hat\r\vert x,p\rangle$
may, up to a well-defined precision, 
be interpreted as the joint probability 
distribution $\r(x,p)$ of the coordinate and momentum. Opposite to 
the coordinate or momentum representations, 
the distribution $\r(x,p)$ contains complete information about the 
system's quantum state and thus satisfies a closed evolution equation [5]. 

In Part 2., I recapitulate the coherent state representation in 
phase space coordinates. 
A most natural statistical interpretation, free of any
reference to measurements or collapses (cf. Ref.~[6]), will be proposed. 
In Part 3., we extend our 
results to the case of composite systems. In the subsequent parts
I stress that typical theories of the emergence of classicality
in quantum theory, including the von Neumann-L\"uders' one, 
could be derived in coherent state representation.

{\it 2. QM in coherent state representation.\/}
Coherent states $\vert x,p\rangle$, labelled by the canonical variables
$x$ and $p$, form a non-orthogonal overcomplet basis in the system's
Hilbert-space. Coherent states are eigenstates of the particular
non-Hermitian combination $\hat x+i\hat p$ of canonical operators:
$$\left(\hat x+ i\hat p\right)\vert x,p\rangle = (x+ip)\vert x,p\rangle.
\eqno(2)$$
The influence of the adjoint operator $\hat x-i\hat p$ brings derivative
terms in:
$$\left(\hat x- i\hat p\right)\vert x,p\rangle = 
  \left(\p_x-i\p_p\right)\vert x,p\rangle
  +{1\over2}(x-ip)\vert x,p\rangle.
\eqno(3)$$
To fix the normalization of the coherent states, we use the
completeness relation in the following form:
$$\int\vert x,p\rangle\langle x,p\vert dxdp = I.
\eqno(4)$$
Then, we consider the diagonal elements of the density operator 
in coherent state representation:
$$\r(x,p)=\langle x,p\vert\hat\r\vert x,p\rangle.
\eqno(5)$$
Obviously, it is a mere technical task to rewrite the abstract 
Schr\"odinger-equation (1) into the above coherent state representation. 
It is,
however, much less obvious that the diagonal part will satisfy a closed
evolution equation [7]:
$${d\over dt}\r(x,p)
=-i\H\left(x+{\p_x+i\p_p\over2},p+{\p_p-i\p_x\over2}\right)\r(x,p) + c.c.
\eqno(6)$$
A few examples are given in the Appendix.

One might think of $\r(x,p)$ as the joint probability distribution of
the phase space coordinates $x$ and $p$ associated with $\hat x$ and 
$\hat p$ or, saying briefly, of the operators
$\hat x$ and $\hat p$ themselves.
The case is, however, 
more complicated because, for all composite variables ${\cal F}(x,p)$, 
one should specify with what Hermitian
operators could they be associated. 
We suggest the following correspondence
rule. Let any classical dynamic variable ${\cal F}(x,p)$ be associated
with the fully symmetrized [8] Hermitian operator 
$\left\{\F(\hat x,\hat p)\right\}_{sym}$. Let us, furthermore,
introduce a Gaussian coarse-graining of the dynamic variables over the 
Planckian phase space cell:
$$
\bar{\F}(x,p)\equiv
\left\langle\F(x+\xi/\sqrt{2},\ p+\eta/\sqrt{2})\right\rangle_{\xi,\eta},
\eqno(7)$$
where $\xi,\eta$ are standard Gaussian variables over which an average 
is understood on the r.h.s. of the above equation.
After these preparations, one can state an equivalence theorem [9].
The quantum expectation value in state $\hat\r$ of a coarse-grained 
fully symmetrized Hermitian dynamic variable
coincides exactly with the classical expectation value of the
corresponding fine-grained
classical dynamic variable, using $\r(x,p)$ for the phase space
probability distribution, i.e.:
$$tr\left(\hat\r\left\{\bar\F(\hat x,\hat p)\right\}_{sym}\right)=
\int\F(x,p)\r(x,p)dxdp
\eqno(8)$$
satisfies for all $\F(x,p)$. 
This is the {\it statistical interpretation} 
which emerges naturally in coherent
state representation. To cut it short, we shall say that $\r(x,p)$ expresses
the coarse-grained joint probability distribution of $\hat x$ and $\hat p$.

{\it 3. QM in hybrid representation.\/}
Let us consider a composite quantum system ${\cal C}\times{\cal Q}$, 
with canonical variables
$\hat x,\hat p$ and $\hat X,\hat P$, respectively.
We introduce a coherent state basis
$\vert x,p\rangle$ for the subsystem ${\cal C}$ while we keep the
abstract Hilbert-space notation for the subsystem ${\cal Q}$. This will be 
called {\it hybrid representation} [9]. Accordingly, by projecting
the
density operator $\hat\r$ of the composite system
${\cal C}\times{\cal Q}$, we introduce the {\it hybrid density} [9]
(cf. also [10,11])
$$
\hat\r(x,p)
\equiv tr\Bigl( (\vert x,p\rangle\langle x,p\vert\otimes I)\hat\r\Bigr).
\eqno(9)$$
This mathematical object might seem strange for the first sight. It
characterizes the quantum state of the original composite system while,
formally, it is a phase space distribution for the subsystem ${\cal C}$
and density operator for the subsystem ${\cal Q}$. 

Similarly to the case in the previous Part, a closed evolution
equation exists for the hybrid density, too [12]: 
$$
{d\over dt}\hat\r(x,p)=
-i\hat\H\left(x+{\p_x+i\p_p\over2},p+{\p_p-i\p_x\over2}\right)
\hat\r(x,p) + h.c.
\eqno(10)$$
where the compact notation ${\cal H}(\hat x,\hat X,\hat p,\hat P)=
\hat{\cal H}(\hat x,\hat p)$ has been applied to the composite 
system's Hamiltonian.

The hybrid density has a rich structure to interpret statistically.
First of all,
$$\hat\r_{{\cal Q}}\equiv \int\hat\r(x,p)dxdp
\eqno(11)$$
is the reduced density operator of subsystem ${\cal Q}$.
Alternatively,
$$\r_{{\cal C}}(x,p)\equiv tr\hat\r(x,p)
\eqno(12)$$
is the coarse-grained 
phase-space distribution of $\hat x,\hat p$ of subsystem ${\cal C}$,
in the very sense of the statistical interpretation of the previous
Part. And there is an additional possibility, common in statistics, 
that is to introduce
$$\hat\r_{{\cal Q}xp}\equiv {\hat\r(x,p)\over\r_{\cal C}(x,p)}
\eqno(13)$$
as the {\it conditional quantum state\/} [9] of the subsystem ${\cal Q}$ 
when the subsystem 
${\cal C}$'s coordinates take given values $x,p$.
This interpretation makes the so-called quantum measurement and
collapse simple consequences of the standard quantum theory as we shall
see in the forthcoming Part.

{\it 4. Application: Collapse.\/}
Consider von Neumann's dynamic model [1] of quantum measurement. 
${\cal Q}$ will denote the measured system and ${\cal C}$
will denote the measuring apparatus. Both of them are  
quantum systems hence we apply standard quantum mechanics to the 
composite system ${\cal Q}\times{\cal C}$. We do so in hybrid 
representation.
For simplicity, the measured system ${\cal Q}$ is a two-state system
initially in the superposition:
$\vert\psi_{{\cal Q}i}\rangle=\alpha\vert0\rangle+\beta\vert1\rangle$.
The apparatus ${\cal C}$ is a canonical quantum system; let the pointer's 
position be $\hat x$. In coherent state representation, ${\cal C}$'s  
initial state $\r_{{\cal C}i}(x,p)$ is concentrated at 
pointer position $x\approx0$ with precision $\Delta x\approx 1$.
Then the initial state of the composite system
${\cal Q}\times{\cal C}$ has the following hybrid density (9):
$$\hat\r_i(x,p)=\vert\psi_{{\cal Q}i}\rangle\langle\psi_{{\cal Q}i}\vert
\r_{{\cal C}i}(x,p).
\eqno(14)$$
According to von Neumann, the interaction during the act of measurement
can be modelled by the interaction Hamiltonian 
$$\hat\H(x,p)=-g\delta(t)\vert1\rangle\langle1\rangle p, 
\eqno(15)$$
where, again, we use hybrid representation. 
The coupling constant $g$ satisfies $g\gg1$.
To calculate the final
entangled state of the composite system, we apply the hybrid equation of
motion (10) with the above interaction Hamiltonian: 
$${d\over dt}\hat\r(x,p)=-g\delta(t)\vert1\rangle\langle1\rangle 
\p_x\hat\r(x,p).
\eqno(16)$$
Then the initial state (14) evolves into a calculable 
final state which, after integrating out the irrelevant momentum
variable $p$, takes the form:
$$\hat\r_f(x)
=\ \vert\alpha\vert^2\vert0\rangle\langle0\vert\r_{{\cal C}i}(x)
+\vert\beta \vert^2\vert1\rangle\langle1\vert\r_{{\cal C}i}(x-g)
$$$$
+\left(\alpha^\star\beta\vert1\rangle\langle0\vert+
    \alpha\beta^\star\vert0\rangle\langle1\vert\right)\r_{{\cal C}i}(x-g/2).
\eqno(17)$$
Let us invoke the statistical interpretation proposed in Part 3.: 
the coarse-grained probability distribution (12) of the pointer's position
$\hat x$ in the final state (17) is 
$$\r_{{\cal C}f}(x)\equiv tr\hat\r_f(x)
=\vert\alpha\vert^2\r_{{\cal C}i}(x)
+\vert\beta\vert^2\r_{{\cal C}i}(x-g),
\eqno(18)$$ 
i.e. $x\approx0$ with probability $\vert\alpha\vert^2$ and $x\approx g$
with probability $\vert\beta\vert^2$. The two outcomes separate well since
we assumed $g\gg1$. The corresponding conditional final quantum
states (13) $\hat\r_{xf}\equiv\hat\r_f(x)/\r_{{\cal C}f}(x)$ are
$\vert0\rangle\langle0\vert$ if $x\approx0$ and 
$\vert1\rangle\langle1\vert$ if $x\approx g$, respectively.
This is the von Neumann-L\"uders collapse [1,2].

{\it 5. Application: Markovian open systems.\/} 
Our example will be the composite quantum system ${\cal Q}\times{\cal C}$,
where ${\cal Q}$ is a two-state atom interacting with the infinite number
of electromagnetic field oscillators ${\cal C}$ (see, e.g., in Ref.[4]). 
For each eigenfrequency $\o$, the canonical oscillator variables 
$\hat x_\o,\hat p_\o$ are, for convenience, 
expressed by the usual non-Hermitian ones $\hat a_\o,\hat a_\o^\dagger$.
If $\vert\psi_{{\cal Q}i}\rangle$
is the atomic initial state then, in hybrid representation, the composite
system's initial state (9) takes the form
$$
\hat\r_i(a,a^\star)
=\vert\psi_{{\cal Q}i}\rangle\langle\psi_{{\cal Q}i}\vert
\exp(-\vert a\vert^2),
\eqno(19)$$
assuming vacuum state for the radiation field initially. 
The interaction Hamiltonian is,
in interaction picture, modelled by
$$\hat\H(a,a^\star)=
\sum_\o g_\o e^{-i(\o-\o_0)t}\hat\s_{+} a_\o + h.c.
\equiv \Gamma\hat\s_{+}+h.c.
\eqno(20)$$
where $g_\o$ are coupling constants, 
$\hat\s_{+}=\vert1\rangle\langle0\vert$,
and $\o_0$ is the atomic energy split. (The time dependence of 
$\hat\H(a,a^\star)$ is suppressed in notation.) 
For later convenience, we have introduced the (complex) effective
field $\Gamma$ to which the atom is really coupled.
In interaction picture, the hybrid equation of motion (10) reads:
$${d\over dt}\hat\r(a,a^\star)=-i[\hat\H(a,a^\star),\hat\r(a,a^\star)]
-i\sum_\o 
\Bigl(
g_\o e^{-i(\o-\o_0)t}\hat\s_{+}{\p\hat\r(a,a^\star)\over\p a^\star} 
- h.c.\Bigr).
\eqno(21)$$

It is well-known that in Markovian approximation the atomic reduced
density operator satisfies the master equation (see, e.g., in Ref.[4])
$$
{d\over dt}\hat\r_{\cal Q}
=\gamma\left( \hat\s_{-}\hat\r_{\cal Q}\hat\s_{+}   
-{1\over2}\hat\s_{+}\hat\s_{-}\hat\r_{\cal Q}
-{1\over2}\hat\r_{\cal Q}\hat\s_{+}\hat\s_{-}\right)
\equiv{\cal L}\hat\r_{\cal Q}\ ,
\eqno(22)$$             
where $\gamma=\r_{\o_0}^2\vert\Gamma_{\o_0}\vert^2$; the spectral density
$\r_\o$ is defined by $\sum_\o\rightarrow\int \r_\o d\o/2\pi$.
Consistent with this reduced dynamics, the Markovian limit of the
composite system dynamics exists as well.
The hybrid equation of motion (21) 
leads to  coupled Wiener (diffusion) processes for the
conditional pure state 
$\hat\r_\Gamma=\vert\psi_\Gamma\rangle\langle\psi_\Gamma\vert$ 
of the atom and the effective field strength $\Gamma$, described 
respectively by the
following Ito-Langevin-equations:
$$
{d\over dt}\hat\r_\Gamma={\cal L}\hat\r_\Gamma+\sqrt{\gamma}
\left(\dot\xi_t^\star(\hat\s_{-}-\langle\hat\s_{-}\rangle)
\hat\r_\Gamma+h.c.\right),
\eqno(23)$$
$$
i\Gamma=\gamma \langle\hat\s_{-}\rangle+\sqrt{\gamma}\dot\xi_t
\eqno(24)$$
where $\langle\hat\s_{-}\rangle=tr(\hat\s_{-}\hat\r_\Gamma)$ and
$\xi_t$ is the standard complex white-noise [13].

Eq.(23) is the heuristic equation of the quantum state diffusion model
[14],
obtained originally to describe the wave function of individual quantum 
systems during their interactions with measuring apparatuses [15]. 

{\it 6. Application: Coupling quantum to classical.\/} 
This time we make the crucial observation that the hybrid equation of
motion (10) could formally be taken as the equation of motion for the
composite system ${\cal Q}\times{\cal C}$ where ${\cal Q}$ is a quantum
system with canonical coordinates $\hat X,\hat P$ but the other subsystem 
${\cal C}$ is genuine classical with coordinates $x$ and $p$.
So, the equation (10) could in principle describe the interaction of a 
true quantum and a true classical subsystem. The shorthand notation
$\H(x,p,\hat X,\hat P)\equiv\hat\H(x,p)$ will be understood again.

To illustrate things, let ${\cal C}$ be a classical harmonic oscillator 
with Hamilton-function $(p^2+x^2)/2$, coupled linearly to a certain 
quantum system ${\cal Q}$ with Hamiltonian $\hat H$.
The total Hamiltonian will be the following:
$$\hat\H(x,p)=\hat H+{p^2+x^2\over2}+\hat J x,
\eqno(25)$$
where $\hat J$ is a certain Hermitian operator for ${\cal Q}$.
For our purpose it will be useful to represent the state of the 
composite system
by the classical phase space density (12) of ${\cal C}$ and 
by the conditional quantum state (13) of ${\cal Q}$. 
Using the Hamiltonian (25), the hybrid equation of motion (10)
leads to a couple of equations for $\r_{\cal C}(x,p)$ and 
$\hat\r_{{\cal Q}xp}$.
The first equation reads:
$$
{d\over dt}\r_{\cal C}(x,p)= 
(x\p_p-p\p_x)\r_{\cal C}(x,p)+\p_p\Jxp\r_{\cal C}(x,p)
\eqno(26)$$
where $\Jxp=tr(\hat J\hat\r_{{\cal Q}xp})$.
This is a classical Liouville-equation. The phase space distribution changes
as if the Hamilton function were the quantum expectation value of
the hybrid Hamiltonian (25):
$$\dot x=\p_p\langle\hat\H(x,p)\rangle_{xp},~~~~
  \dot p=-\p_x\langle\hat\H(x,p)\rangle_{xp}.
\eqno(27)$$
These equations are identical to the fenomenological mean-field equations 
widely used to calculate the evolution of a classical
system under the influence of a quantized one (see, e.g., in Ref.[11]). 
The second equation will
govern the quantized system's state under the influence
of the classical subsystem.  
Our hybrid equation of motion (10) yields, after some
elaboration, the following form:
$$
{d\over dt}\hat\r_{{\cal Q}xp}+(\dot x\p_x +\dot p\p_p) \hat\r_{{\cal Q}xp}
+i[\hat\H(x,p),\hat\r_{{\cal Q}xp}]=
$$$$
=-{i\over2}(\p_x s)[\hat J,\hat\r_{{\cal Q}xp}]
-{i\over2}[\hat J,\p_x\hat\r_{{\cal Q}xp}]
+{1\over2}\p_p\{\hat J-\Jxp,\hat\r_{{\cal Q}xp}\}
+{1\over2}(\p_p s)\{\hat J-\Jxp,\hat\r_{{\cal Q}xp}\}
\eqno(28)$$
where
$s(x,p)\equiv\ln\r_{\cal C}(x,p)$. This equation differs from the 
corresponding equation of the standard mean-field theory
by the terms on the r.h.s which are absent in the
fenomenological theory. The standard theory can not describe the
stochastic fluctuations induced in the classical subsystem ${\cal C}$ by
the quantum subsystem ${\cal Q}$. The terms on the r.h.s. take these
fluctuations into the account [16].

{\it Moral.\/}
The classical content, if there is any, of a given quantum state displays
itself in coherent state representation. I tend to believe that this is
the general case for canonical quantum systems. 
It follows and is important to realize that the classical content is a 
relative and apparent feature. We can read it out if we use the proper 
framework --- the coherent state representation, presumably.
On the other hand, such classicality might be a robust feature
against changes of the framework, here I mean canonical transformations on 
the first place. 
Obviously, the coherent state representation is not invariant
canonically. But, due to the well-defined amount of coarse-graining used
in the statistical interpretation, one hopes a satisfactory robustness of
classicality whenever the dynamic variables $\F(x,p)$ have no structure
at short scales $\Delta x\approx\Delta p\approx1$.

It is rather promizing that the concept, presented in this
talk, seems to cover all alternative concepts, including 
the wave function collapse, the quantum state diffusion, 
and the interpretation based on true classical subsystems [17]. 
I think that a Bohm-type interpretation, too, follows naturally
(cf., maybe, Ref.~[18]). Though I did not mention the decoherent
history interpretation I have no doubts that hybrid quantum dynamics
turn out to build up decoherent histories (we do not need 
the Markovian limit [19] either).
\bigskip

I am deeply indebted to the organizers of the Symposium for inviting me
to participate and to give this talk.
This work was supported by the Hungarian Scientific Research
Fund under Grant OTKA T016047.
\parindent=0in
{\bf APPENDIX}
\bigskip
Harmonic oscillator, $\H=(p^2+x^2)/2$:
$${d\over dt} \r(x,p)=(x\p_p-p\p_x) \r(x,p).$$
Free particle, $\H=p^2/2$:
$${d\over dt} \r(x,p)=-p\p_x \r(x,p) +{1\over2}\p_x\p_p \r(x,p).$$
Quartic oscillator, $\H=(p^2+x^2)/2+\lambda x^4$:
$$
{d\over dt} \r(x,p)=(x\p_p-p\p_x) \r(x,p)
-\lambda\left(x+{\p_x\over2}\right)^3\p_p\r(x,p)
$$$$
+{\lambda\over4}\left(x+{\p_x\over2}\right)\p_p^3\r(x,p).
$$
\bigskip

{\bf REFERENCES and FOOTNOTES}\hfill
\bigskip
1. J.von Neumann, {\it Matematische Grundlagen der Quanten Mechanik}
        (Springer, Berlin, 1932).\hfill\break
2. G.L\"uders, Ann.Phys. (Leipzig) {\bf 8}, 322 (1951).\hfill\break
3. V.Bargmann, Commun.Pure Appl.Math. {\bf 14}, 187 (1961);
   R.J.Glauber, Phys.Rev. {\bf 131}, 2766 (1963); see also in
   the recent textbook [4].\hfill\break
4. D.F.Walls and G.J.Milburn, {\it Quantum Optics}
        (Springer, Berlin, 1994).\hfill\break
5. A.Lonke, J.Math.Phys. {\bf 19}, 1110 (1978).\hfill\break
6. J.S.Bell, Phys.World {\bf 3}, 33 (1990).\hfill\break 
7. This equation preserves the positivity of $\r(x,p)$ provided certain
   analiticity conditions are satisfied initially. It preserves the
   pure state property of $\hat\r$ though all forthcoming formulae
   will be valid for mixtures as well. The proofs will be given elsewhere. 
   The eq.~(10) has an equivalent compact form:
$${d\over dt}\r
=\H\ 2\sin{\pl_x\pr_p-\pl_p\pr_x\over2}
      \exp{\pl_x\pr_x+\pl_p\pr_p\over2}\ \r
=\{\H,\r\}_{Poisson}+h.o.t.
$$
On the very right, one may observe the classical Liouville evolution
equation to appear in the lowest order of the derivatives.\hfill\break
8. Full symmetrization (Weyl-ordering) is defined by the recursive 
rules
$$\left\{\hat x\F(\hat x,\hat p)\right\}_{sym}=
 {1\over2}\left\{\hat x,\left\{\F(\hat x,\hat p)\right\}_{sym}\right\}
,\ \ \ \ 
  \left\{\hat p\F(\hat x,\hat p)\right\}_{sym}=
 {1\over2}\left\{\hat p,\left\{\F(\hat x,\hat p)\right\}_{sym}\right\},
$$
while
$\{\hat x\}_{sym}=\hat x$ and $\{\hat p\}_{sym}=\hat p$.\hfill\break
9. L.Di\'osi, Quant.Semiclass.Opt. {\bf 8}, 309 (1996); 
   L.Di\'osi, "A true equation to couple classical and quantum dynamics",
   e-print archives quant-ph/9610028.\hfill\break
10. I.V.Aleksandrov, Z.Naturf. {\bf 36A}, 902 (1981).\hfill\break
11. W.Boucher and J.Traschen, Phys.Rev. D {\bf 37}, 3522
   (1988).\hfill\break
12. Expansion in the derivatives yields the form
$$
{d\over dt}\hat\r=-i[\hat\H,\hat\r]
             +{1\over2}\{\hat\H,\hat\r\}_{Poisson}
             -{1\over2}\{\hat\r,\hat\H\}_{Poisson}
$$$$             
\ \ \ \ \ \ \ 
             -{i\over2}[\p_x\hat\H,\p_x\hat\r]
             -{i\over2}[\p_p\hat\H,\p_p\hat\r]+h.o.t.
$$
showing the tricky combination of Liouville's classical and Schr\"odinger's
quantum evolutions. The first line of the r.h.s. is identical to the
Aleksandrov-bracket [10] which, in itself, would violate the positivity
of $\hat\r$ [11].\hfill\break
13. Proofs will be given elsewhere.\hfill\break
14. N.Gisin and I.C.Percival, J.Phys. A {\bf 25}, 5677 (1992).\hfill\break 
15. L.Di\'osi, J.Phys. A {\bf 21}, 2885 (1988).\hfill\break 
16. Obviously, if the quantum fluctuation of the "current" $\hat J$ is 
    small enough then the last two r.h.s. terms go away. If, furthermore, 
    the states $\r_{\cal C}(x,p)$ and $\hat\r_{{\cal Q}xp}$ are smooth 
    enough functions of $x,p$ then all r.h.s. terms can be ignored and we 
    are left with the standard mean-field equation
$$
{d\over dt}\hat\r_{{\cal Q}xp}+i[\hat\H(x,p),\hat\r_{{\cal Q}xp}]=0.
$$
17. T.N.Sherry and E.C.G.Sudarshan, Phys.Rev. {\bf D18}, 4580 (1978).
    \hfill\break
18. G.G. de Polavieja, Phys.Lett. {\bf 220A}, 303 (1996).\hfill\break
19. L.Di\'osi, N.Gisin, J.Halliwell, and I.C.Percival, Phys.Rev.Lett.
    {\bf 74}, 203 (1995).\hfill\break
\end